\documentclass{article}
\usepackage{graphicx}
\usepackage[english]{babel}
\usepackage{amsmath}
\usepackage{multicol}
\usepackage{graphicx}
\usepackage{tabularx}
\usepackage{epigraph}
\usepackage{lineno}
\usepackage[numbers,sort&compress]{natbib}

\usepackage[margin=2cm]{geometry}

\title{Christmas tree-like Coulomb crystals}
\author{Rudyi S.S.,  Romanova A.V., Tatarinov D.A., \\ Kurdyukov D.A., Stovpiaga E. Yu., Golubev V.G., and Shcherbinin D.P.}

\begin{document}
\maketitle
\begin{abstract}
We have investigated the formation of microparticles-based Coulomb crystals in a linear vertical quadrupole Paul trap with a single end-cap electrode. For the first time, we have described the effect of gravity on Coulomb crystals formation. We have observed a new type of stable Coulomb crystals configuration that we referred to as ``Christmas tree-like Coulomb crystals''. We have provided numerical simulation and experimental research of the ``Christmas tree-like Coulomb crystals'' and discussed these structures in the perspective of Science and Art performance.
\end{abstract}
\section{Introduction}
\epigraph{Oh, beauty! Beauty is essential. You can easily do without any theories or any experimental studies proving them. You can somehow live without conferences or publications in highly ranking journals. You may not believe it, but there was once a little boy who found a way to even do without quantum mechanics. But you cannot possibly survive without beauty.
 \\No way!}{\textit{D.P.}}

Can you imagine Christmas without a Christmas tree? For many people, the Christmas tree is a symbol of magic. Sadly, many of us claim that there is no room for magic in this rational world. However, this is not about magic or miracles, this is about ourselves.  We are always in a hurry, unaware of anything around us. We are lost in the crowd commuting on a frosty morning, and are blind to the rays of the rising sun breaking up into thousands of glittering reflections on the snow. We hurry back home and don't feel the barely perceptible smell of tangerines emanated by a person next to you on the metro.  In fact, there is magic everywhere. Stop for a moment and take a closer look!  Miracles happen even where you least expect them. You may not believe it, but a miracle did happen in a small laboratory in a majestic septentrional city. And here is a true story about the tiny particles circling in an intricate dance, cheerfully obeying their strict choreographer - His Majesty the Electric Field. In our routine lab work we would say, ``That's a Coulomb crystal in a Paul trap''. But on Christmas eve, magic is everywhere. Reality and fairy tale come together, and a common Coulomb crystal becomes a fascinating Christmas tree, lit by hundreds of perovskite lights.

Well! At this point, let us to turn to physics. This article focuses on a special form of cold plasma matter known as Coulomb Crystals~\cite{thompson2015ion,willitsch2012coulomb,mihalcea2023physics}. Coulomb Crystals (CCs) are an ensemble of Coulomb-interacting charged particles. One of the most popular tools to prepare, manipulate and study CCs are linear quadrupole Paul traps~\cite{willitsch2012coulomb,drewsen2015ion,okada2010characterization}. Individual charged particles get trapped in a Paul trap due to their interaction with the fast-oscillated electric field of the trap. With many-body localisation, both particle-field and particle-particle interactions affect CCs formation. In fact, other forces, both electrical and non-electrical nature, can influence the formation of CCs~\cite{mihalcea2023physics,shcherbinin2023charged,laupretre2019controlling,d2021radial}.

A well studied example of Coulomb crystals is CCs from atomic ions in Paul traps. The shape of these CCs is known to depend on the trap geometry, the power supply parameters and the number of ions~\cite{drewsen2015ion,hornekaer2001structural, okada2010characterization,thompson2015ion}. It is common to identify the following stable configurations of atomic ion CCs: one dimensional (1D) CCs including chain crystals and 1D-overcompressed crystals~\cite{drewsen2015ion,romanova2023one}; two dimensional (2D) CCs including ``zig-zag'' and radial crystals~\cite{d2021radial, drewsen2015ion, dubin1993theory, mielenz2013trapping, pyka2013topological}; three dimensional (3D) CCs including spiral-structure crystals and ``Old-fashioned Chandelier quasi-Coulomb Crystals'', to name a few~\cite{dubin1993theory,rudyi2023fractal, horak2012optically, ludwig2005structure, herskind2009realization}.

The ion based CCs have found their application as a promising platform for quantum computing~\cite{thompson2015ion,pogorelov2021compact}. On the other hand, ionic CCs are widely used to detect high energy particles and dark ions~\cite{clark2010detection,schmid2022number}. And, of course, trapped ions and Coulomb crystals can attract both researchers and inquisitive lay public as science art (S\&A) object. Indeed, the photo of a single trapped ion has won the science photography prize in EPSRC contest~\cite{Zachos_2021}. We believe that the convergence of science and art may find more than one application. Firstly, science art (S\&A) is a way to  popularize contemporary science. Secondly, researchers use S\&A approaches in their everyday work. It contributes to their creative thinking and other essential soft skills, which are crucial for new breakthrough ideas. Thirdly, continuously referring to S\&A may prevent researchers from being emotionally exhausted.  

Well, shall we go back to Coulomb Crystals formation process? Although the above discussion is related to atomic ion based CCs, there is no fundamental restriction to generalise the CCs concept to charged nano- and microparticles. On the other hand, the  formation of CCs consisting of microparticles (MCCs) does not seem to have been much explored. The  types of CCs stable configurations have not been systematically classified yet. To the best of our knowledge, no studies have considered the gravity effect on CCs formation. Moreover, the large micromotion amplitude, which highly exceeds the particle size, and which is an essential feature of microparticle trapping process, has not receive any attention either. And of course, MCCs give new opportunities for creating S\&A objects for art performances and exhibitions.

Here, we investigate the formation of microparticles-based CCs in linear vertical quadrupole Paul trap with a single end-cap electrode. For the first time, we describe the effect of gravity on CCs formation. We observe a new type of stable CCs configuration referred to as ``Christmas tree-like Coulomb crystals''. We develop a theoretical model for comprehensive studies of processes within many-body localization of charged microparticles in quadrupole Paul trap. We describe the outcomes of the ``Christmas tree-like Coulomb crystals'' experimental research. We discuss these structures in the S\&A perspective. For the reader's pleasure and on the eve of the glowing Christmas, we have prepared Christmas tree-like Coulomb crystals consisting of hybrid microstructures with luminescent perovskite nanocrystals (pNCs).

\section{Theory of Microparticles-based Coulomb crystals}
For a comprehensive description of MCCs formation the following forces are to be considered: electrical interaction of each particle with the electric field of the trap, the interparticle Coulomb interaction, the dissipative forces, and particle gravity. Thus, the dynamics of each individual particle from CCs can be described by the following equations: 
\begin{equation}
    m\ddot{x}=-e\frac{\partial U}{\partial x}-e\frac{\partial C}{\partial x}-\xi(\dot{x},\dot{y},\dot{z}),\quad   
    m\ddot{y}=-e\frac{\partial U}{\partial y}-e\frac{\partial C}{\partial y}-\xi(\dot{x},\dot{y},\dot{z}),\quad     
    m\ddot{z}=-e\frac{\partial U}{\partial z}-e\frac{\partial C}{\partial z}-\xi(\dot{x},\dot{y},\dot{z})-m g, 
\end{equation}
where $\{x,y,z\}$ are the Cartesian coordinates of each particle, $m$ is the particles mass, $e$ is the particles charge, $U$ is the electric potential distribution, $C$ is the Coulomb interaction, $\xi(\dot{x},\dot{y},\dot{z})$ is the energy dissipation force. 

\begin{figure}
\centering
    \begin{minipage}{0.7\textwidth}
        \includegraphics[width=\textwidth]{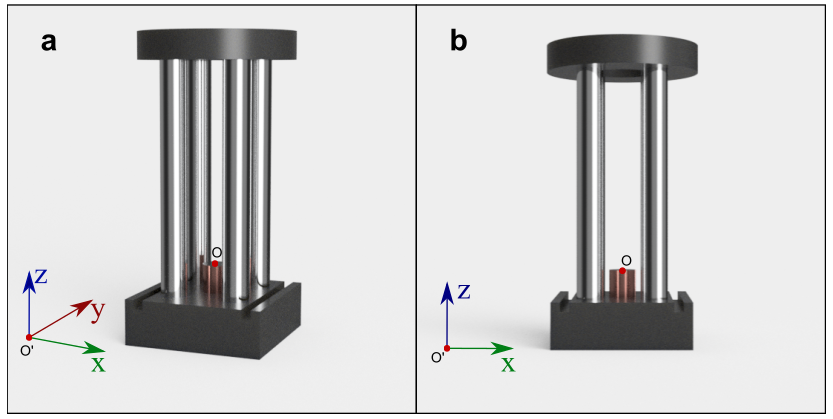}
    \end{minipage}
\caption{\label{fig:cup-trap} Vertical linear quadrupole trap with a single end-cap electrode.}
\end{figure}

The electric field distribution inside the trap is determined by the trap geometry. In the present research, we deal with a ``cup-trap'' which is a vertically oriented linear quadrupole Paul trap with one end-cap electrode. Figure~\ref{fig:cup-trap} is a representation of the ``cup-trap''. The trap vertical electrodes are oriented along $z$-axis. The origin lies in the center of end-cap electrode. This configuration is a prospective solution for localizing charged microparticles. Indeed, as there is no upper electrode, particles can be injected ``from the top'', whereas massive charged particles are retained along the $z$-axis by the counteraction of the particle gravity force and the electrostatic repulsion from the end-cap electrode. 

In a ``cup-trap'' with a disc-shaped end-cap electrode, the electric potential distribution takes the following form
\begin{equation}
    U(x,y,z)\approx\frac{V \cos{(\omega t)}}{2r_0^2}(x^2-y^2)+U_\mathrm{end}(x,y,z),
\end{equation}
where $V$ is the AC voltage amplitude, $\omega$ is the AC voltage frequency, $r_0$ is the trap radius, $U_\mathrm{end}(x,y,z)$ is the potential distribution along the end-cap electrode. For a disc-shaped electrode with the radius $r_E$, the electric potential $U_\mathrm{end}(x,y,z)$ can be written as~\cite{rudyi2020outside}
\begin{equation}
    U_{\mathrm{end}}(x,y,z)=V_\mathrm{0}\left[1-f_\mathrm{E}\right],\label{eq:uend}
\end{equation}
where
\begin{equation}
   f_E=\frac{z \left[\frac{\left(\sqrt{R^2+z^2}-r_E\right) \Pi^+}{\sqrt{R^2+z^2}+R}+\frac{\left(r_E+\sqrt{R^2+z^2}\right) \Pi^-}{\sqrt{R^2+z^2}-R}\right]}{\pi  \sqrt{\left(r_E+R\right){}^2+z^2}} 
\end{equation}
and
\begin{eqnarray}
  \Pi^\pm=\Pi \left(\frac{\pm2 R}{\sqrt{R^2+z^2}\pm R}|\frac{4 R\cdot r_E}{\left(r_E+R\right){}^2+z^2}\right)  
\end{eqnarray}
$\Pi$ is the elliptic Integral of the Third Kind, $R$ is the radius vector in the $xy$-plane $R=\sqrt{x^2+y^2}$, $V_0$ is the DC voltage on the end-cap electrode. 

In this study we consider the CCs consisting of $N$ microparticlces. Thus, the Coulomb potential energy of the $i$-th particle can be written as

\begin{equation}
\label{Coulomb}
    C_i=k\sum_{j=1}^{N, i\neq j}{\frac{e_j}{\left[(x_i-x_j)^2+(y_i-y_j)^2+(z_i-z_j)^2\right]}},
\end{equation}

where $k$ is the electrostatic constant.

When microparticles are localized in the air, the Stokes drag force ($D$) brings about energy dissipation. The Stokes drag force is generally written as

\begin{equation}
    D_i=6\pi\mu r_i \nu_i, 
\end{equation}
where $\nu_i$ is the velocity of the $i$-th particle and $r_i$ is the hydrodynamic radius of the $i$-th particle.

Taking into account all forces, we can write the equations of motions for the $i$-th particle as

\begin{eqnarray}
    m_i\ddot{x_i}=-\frac{e_iV \cos{(\omega t)}}{r_0^2}x_i+\frac{e_i V_0\partial f_\mathrm{E}}{\partial x_i}-ke_i\sum_{j=1}^{N, i\neq j}\frac{e_j(x_i-x_j)}{\left[(x_i-x_j)^2+(y_i-y_j)^2+(z_i-z_j)^2\right]^{3/2}}-6\pi\mu r_i\dot{x_i}
    \\
    m_i\ddot{y_i}=\frac{e_iV \cos{(\omega t)}}{r_0^2}y_i+\frac{e_i V_0\partial f_\mathrm{E}}{\partial y_i}-ke_i\sum_{j=1}^{N, i\neq j}\frac{e_j(y_i-y_j)}{\left[(x_i-x_j)^2+(y_i-y_j)^2+(z_i-z_j)^2\right]^{3/2}}-6\pi\mu r_i\dot{y_i}
    \\
    m_i\ddot{z_i}=\frac{e_i V_0\partial f_\mathrm{E}}{\partial z_i}-ke_i\sum_{j=1}^{N, i\neq j}\frac{e_j(z_i-z_j)}{\left[(x_i-x_j)^2+(y_i-y_j)^2+(z_i-z_j)^2\right]^{3/2}}-6\pi\mu r_i\dot{z_i}-m_i g
\end{eqnarray}

To simplify, we can presume that all particles have similar masses and charges, and the end-cap radius $r_E$  is half of the trap radius, $r_E\to r_0/2$. For further simplifying we can introduce the following mass-spectroscopy like substitutions
\begin{equation}
\label{Sub}
    q=\frac{2eV}{m\omega^2r_0^2}, a=\frac{4eV_0}{m\omega^2r_0^2}, \zeta=\frac{4 k e^2}{m\omega^2r_0^3}, \gamma=\frac{12\pi\mu r}{m\omega}, \alpha = \frac{4g}{\omega^2 r_0},\tau=\frac{\omega t}{2}
\end{equation}

Finally, the reduced equations of motion take the following form

\begin{eqnarray}
\label{MSLsystem}
    \ddot{x_i}=-2q x_i \cos{2 \tau}+a \frac{\partial f_\mathrm{E}}{\partial x_i}-\zeta\sum_{j=1}^{N, i\neq j}\frac{(x_i-x_j)}{\left[(x_i-x_j)^2+(y_i-y_j)^2+(z_i-z_j)^2\right]^{3/2}}-\gamma \dot{x_i}
    \\
\label{MSLsystemy}    \ddot{y_i}=2q y_i \cos{2 \tau}+a \frac{\partial f_\mathrm{E}}{\partial y_i}-\zeta\sum_{j=1}^{N, i\neq j}\frac{(y_i-y_j)}{\left[(x_i-x_j)^2+(y_i-y_j)^2+(z_i-z_j)^2\right]^{3/2}}-\gamma \dot{y_i}
    \\
 \label{MSLsystemz} \ddot{z_i}=-a\frac{\partial f_\mathrm{E}}{\partial z_i}-\zeta\sum_{j=1}^{N, i\neq j}\frac{(z_i-z_j)}{\left[(x_i-x_j)^2+(y_i-y_j)^2+(z_i-z_j)^2\right]^{3/2}}-\gamma \dot{z_i}-\alpha
\end{eqnarray}

To analyze the effect of gravity it is convenient to introduce one more substitution into the equation of motion along the $z$-axis~(\ref{MSLsystemz})
\begin{equation}
    \label{adds}
\eta=\frac{\alpha}{a}=\frac{m g {r_0}}{e V_0},
\end{equation}
 which has a simple physical implication of the relation between the gravity and the electrostatic repulsion interaction.

 Using the latter substitution, the equation~(\ref{MSLsystemz}) can be rewritten as
\begin{equation}
    \label{newz}
 \ddot{z_i}=-a(\frac{\partial f_\mathrm{E}}{\partial z_i}+\eta)-\zeta\sum_{j=1}^{N, i\neq j}\frac{(z_i-z_j)}{\left[(x_i-x_j)^2+(y_i-y_j)^2+(z_i-z_j)^2\right]^{3/2}}-\gamma \dot{z_i}
\end{equation}
 
If the gravity is negligible compared to electric forces ($a\gg\alpha$), $\eta \to 0$. In this case, the equations of motion (\ref{MSLsystem})-(\ref{newz}) corresponds to those for atomic ions. With the increase of the gravity effect the homogeneity terms $\eta$ as well as $\alpha$ become non-negligible.  

\begin{figure}[ht!]
    \centering
    \begin{minipage}{0.48\textwidth}
    \includegraphics[width=0.75\textwidth]{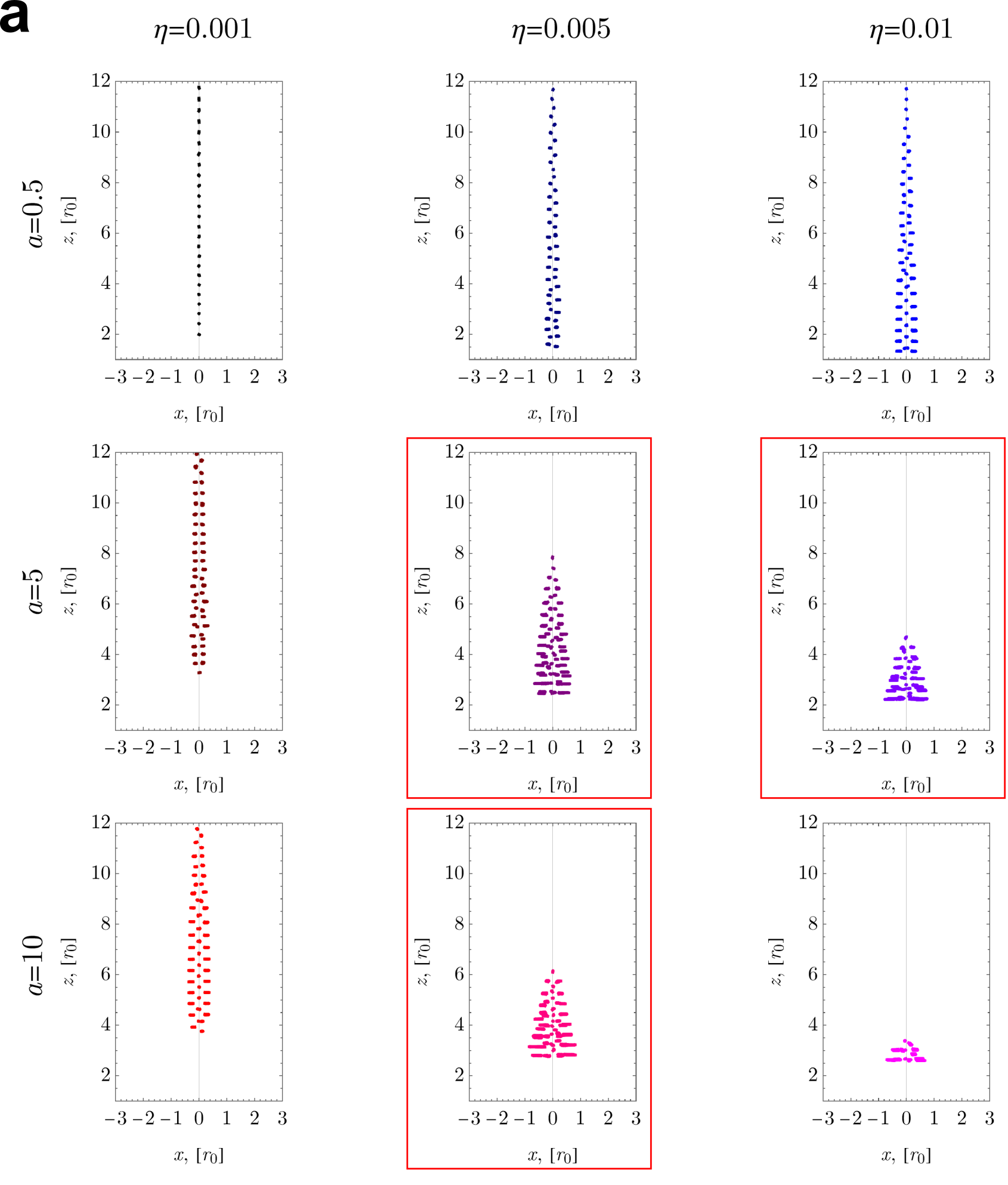}
    \end{minipage}
    \begin{minipage}{0.48\textwidth}
    \includegraphics[width=0.90
    \textwidth]{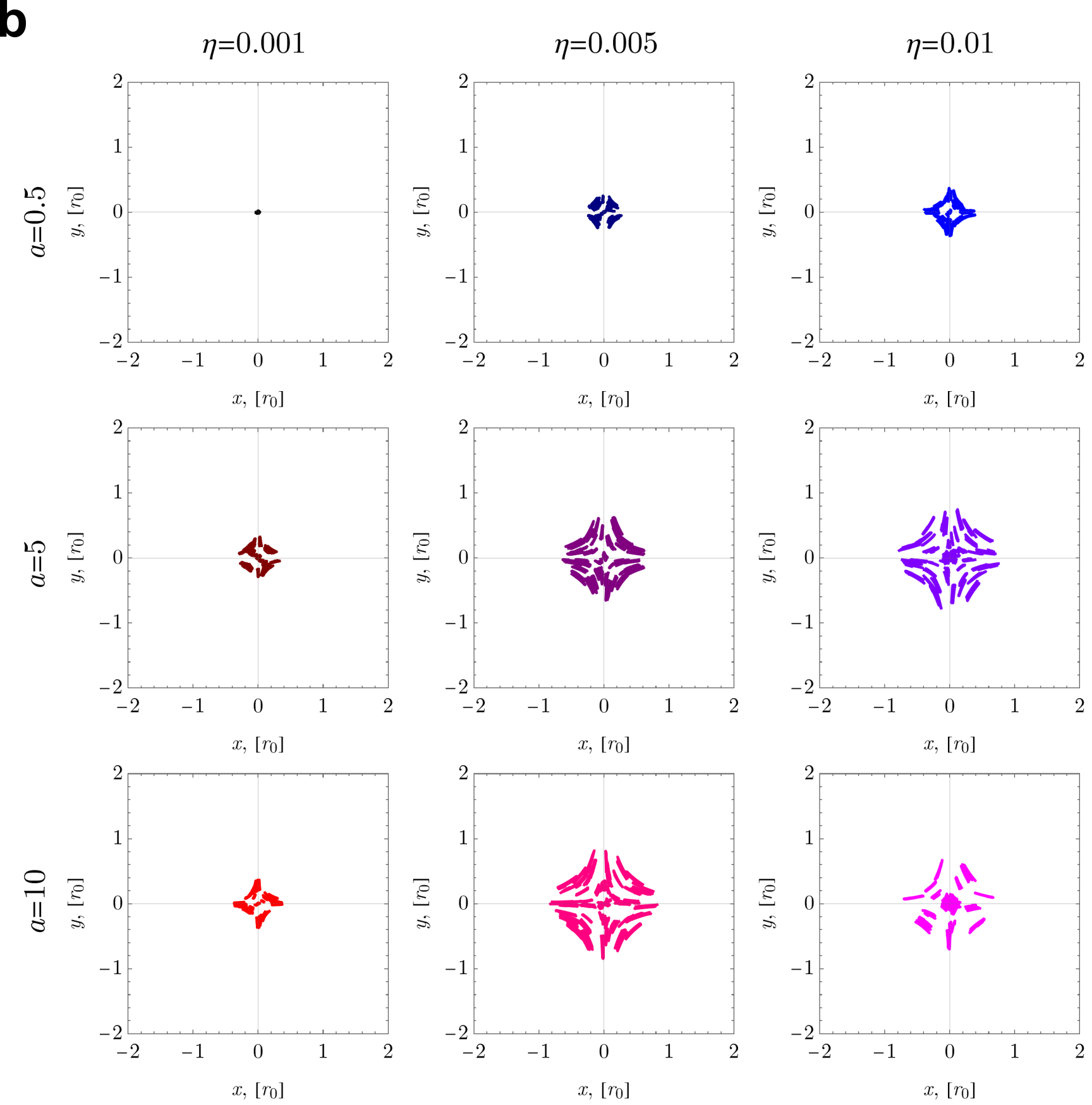}
    \end{minipage}
    \\        
    \caption{Simulation of MCCs dynamics with $a=0.1$ (part a), 0.5 (marked in red, part b), 5 (marked in blue, part c), and 10 (marked in orange, d) for $\eta$ lying in the range $\eta\in[0.001, 0.01]$.}
    \label{fig:sim}
\end{figure}

To study the effect of gravity on CCs formation, we simulated many-body localisation according to (\ref{MSLsystem})-(\ref{newz}) with the fixed values of $q$, $\zeta$ and $\gamma$, and the $a$ and $\eta$ parameters varied. To perform a simulation of the many-body localization for 128 charged particles according to~(\ref{MSLsystemz}), we numerically solved the Cauchy problem using the Runge-Kutta method. The radial initial coordinates were defined randomly in the interval $\{x,y\}\in[-0.5r_0, 0.5r_0]$. The axial initial coordinate was in the interval $\{z\}\in[5r_0, 10r_0]$ which corresponded to the injection ``from the top''. The initial velocities were zero. Figure~\ref{fig:sim} shows the results of the simulation for the parameters $a=0.5, 5, 10$ and $\eta=0.001, 0.005, 0.01$. For each pair of parameters Figure~\ref{fig:sim} provides $\mathrm{XZ}$ and $\mathrm{XY}$ projections corresponding to the front and the top view. Other parameters of dynamical system are given as $q=0.7$, $\xi=0.0015$, $\gamma=0.025$ and fixed $\forall{a,\eta}$. The integration time corresponds to the $\tau_{max}=10^3$.

The results shown on Figure~\ref{fig:sim} prove that CCs can be formed from microparticles. Moreover, the shape and the inner structure of crystals depend on the parameters $a$ and $\eta$.  Figure~\ref{fig:sim} shows that depending on the above parameters, various stable configurations of MCCs can be observed ranging from a linear chain to 3D structures. With the increase of the $\eta$-value for all $a$-values, the MCCs become wider and shorter. This phenomenon can be accounted for by the fact that $\eta$ depends on the ratio of the gravity and the electrostatic repulsion. Thus, the increase of the gravity effect (the higher $\eta$ value according to~\ref{adds}) results in MCCs deformation. For some parameters we can observe MCCs with cone shapes. Such structures are highlighted in red bold frames. They really do look like small Christmas trees, and we dared use this newly coined term to refer to such structures.

To give reasons for the term ``Christmas tree-like MCCs'' we interviewed 35 people from the International research and education center for physics of nanostructures. We asked them to recognize Christmas-tree like structures from a set of $20$ different MCCs shapes. We showed them the MCCs front views which corresponded to some random pairs of $a$ and $\eta$. We assumed that a structure can be referred to as Christmas tree-like, if at least 80\% of respondents claimed the MCCs similarity to a Christmas tree. Our colleagues appeared to recognize Christmas tree-like structures from the whole set when the cone-shape MCCs have the height-to-base diameter ratio in the range of $\sim1-5$. It is worth mentioning that real spruce have the same ratio of the height to the maximal crown width, according to the investigation published in 2023~\cite{ahmed2023tlidar}. In fact, our respondents made decisions rather intuitively, without any specific data related to geometric proportions. Thus, this by-product of our research is very promising in the perspective of cognitive science and visual pattern recognition. We will be happy if someone from cognitive neuroscience or psychology developed our observations to a comprehensive independent study of the human processes of pattern recognition of Coulomb crystals shapes and those of trees, and a correlation between them, if any. 

Shall we come back to physics again? Figure~\ref{fig:sim} demonstrates that the oscillation amplitude of the particles located near the $z$-axis is much lower than that of the particles farther from the axis. In terms of the Christmas tree, the oscillation of the particles forming the tree trunk has almost degenerated, while the needle-based particles oscillate with a high amplitude. To explain this behaviour, we have to analyse the equations~(\ref{MSLsystem}-\ref{MSLsystemy}). The first fast-oscillating terms are proportional to the coordinates $x_i$ and $y_i$. This means that when the particle moves away from the ``tree trunk'', the force exerted by the linear electrodes rises proportionally to the distance. Figure~\ref{fig:sim} shows the complete particle trajectories after MCCs stabilization. Technically, if we try to get a photo of in-lab implemented MCCs with the exposition time higher than $2 \pi/\omega$, we will see the tracks corresponding to particle trajectories. Moreover, the trajectories of the needle-based particles repeat the hyperbolic form of equipotential electric field lines. 

\section{Experimental part} 

\begin{figure}[ht!]
\centering
        \begin{minipage}{0.7\textwidth}
        \includegraphics[width=\textwidth]{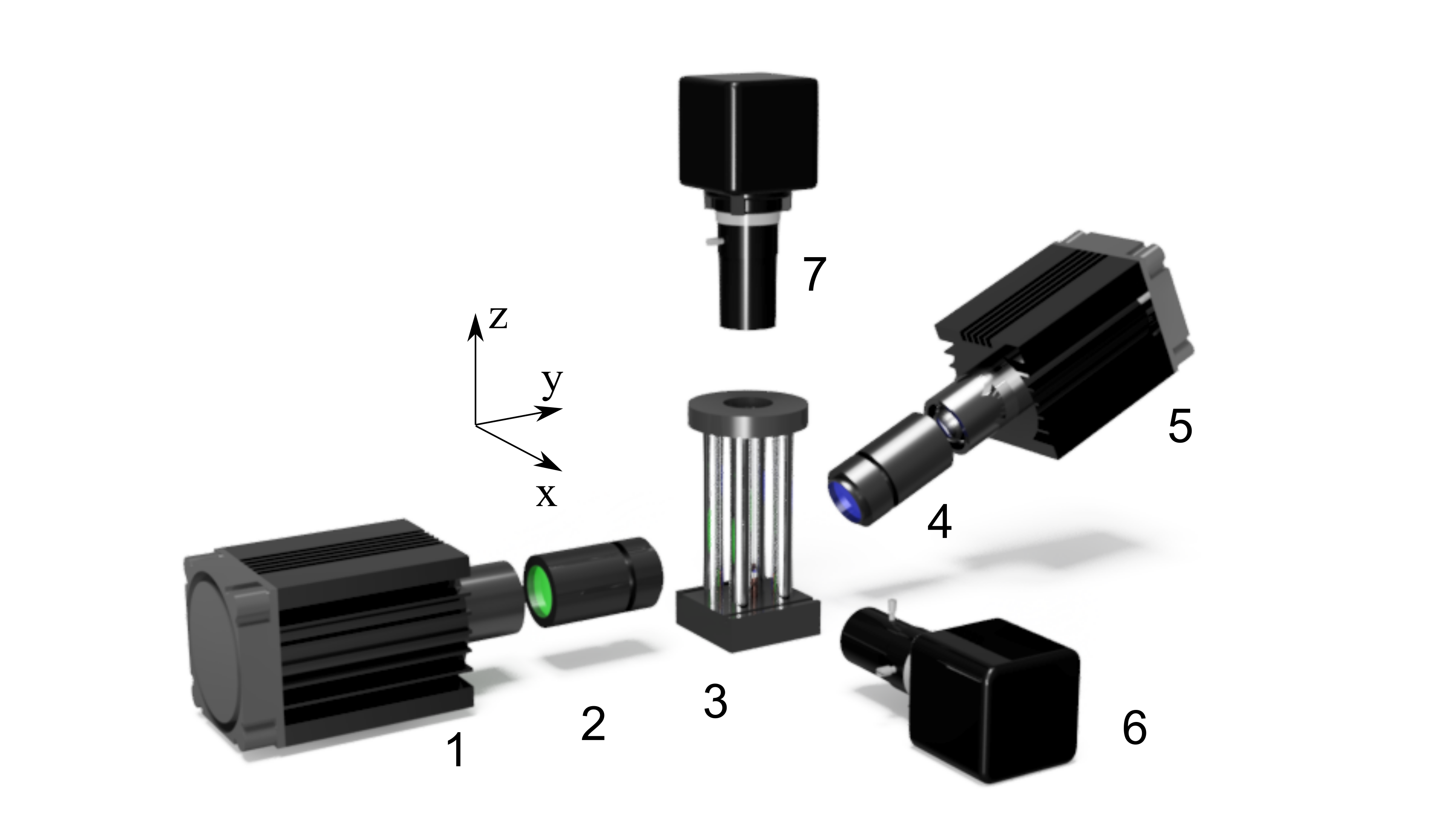}
    \end{minipage}    
    \caption{Scheme of the experimental setup for MCCs trapping. 1 -- 532 nm DPSS laser, 2,4 -- beam expanders, 3 -- quadrupole ion trap, 5 - 375 nm laser, 6,7 - cameras}
    \label{fig_exp:ExperimentalSetup}
\end{figure}

\begin{figure}[ht!]
\centering
    \begin{minipage}{0.25\textwidth}
\centering\includegraphics[width=\textwidth]{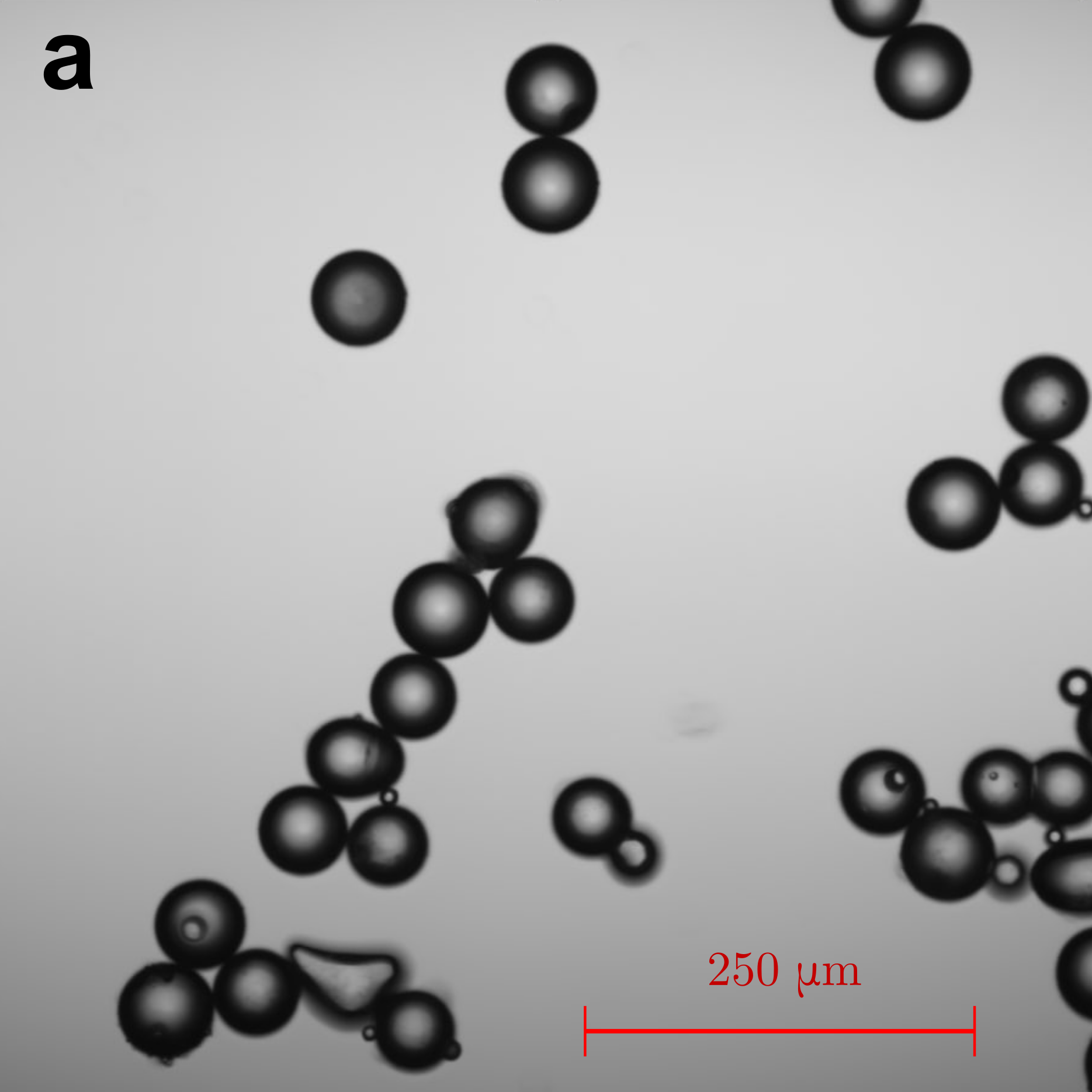}
    \end{minipage}   
    \quad
    \begin{minipage}{0.245\textwidth}
    \centering\includegraphics[width=\textwidth]{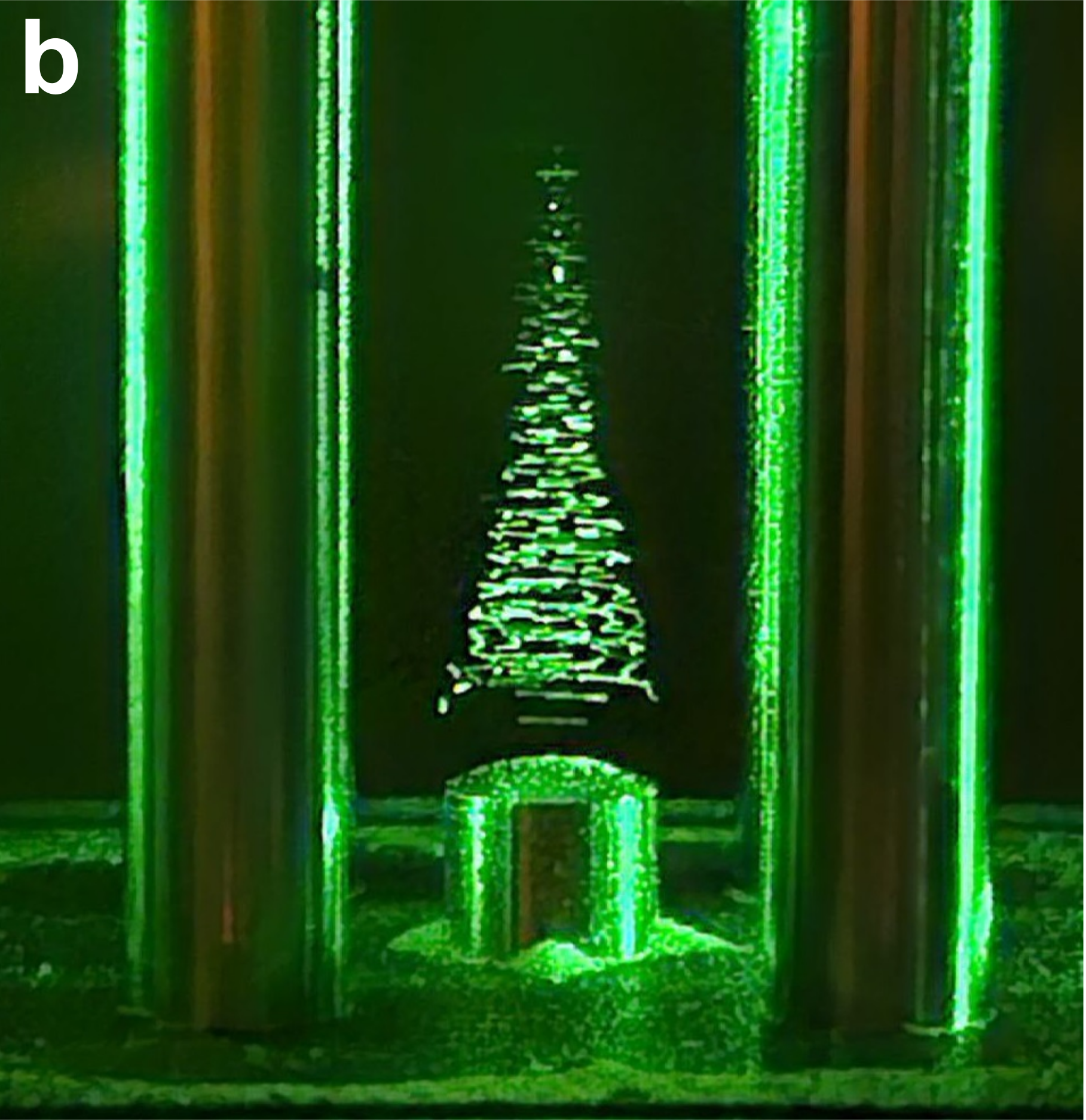}
    \end{minipage}\quad
    \begin{minipage}{0.255\textwidth}
    \centering\includegraphics[width=\textwidth]{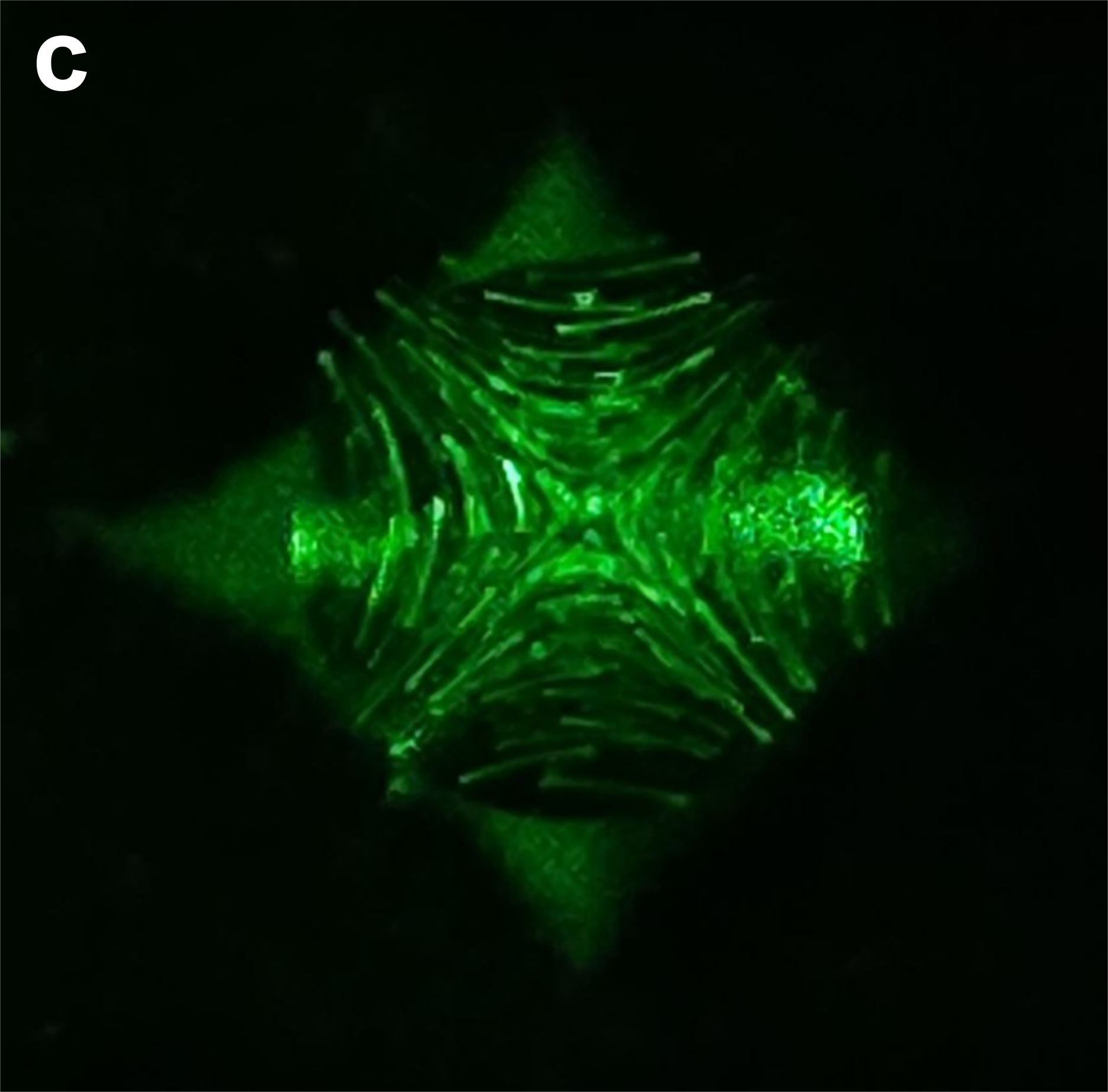}
    \end{minipage}
\caption{\label{fig:greentree} Silica-based MCCs structures; a) -- microscope image of microspheres, b) -- front view of the Christmas tree-like silica-based MCCs, c) -- top view of the Christmas tree-like silica-based MCCs}
\end{figure}

In this section we provide an experimental validation of forming Christmas tree-like MCCs in a ``cup-trap''. Two types of particles were used: the non-luminescent silica microparticles to study the internal structure of Christmas tree-like MCCs and luminescent hybrid structures with perovskite nanoparticles to create the S\&A object based on MCCs. We used a ``cup-trap'' (Figure~\ref{fig:cup-trap}) at atmospheric pressure and room temperature (293K). A scheme of the experimental setup is shown in Figure~\ref{fig_exp:ExperimentalSetup}. To provide radial confinement ($XY$ plane), four linear electrodes were diagonally paired and connected to an AC power supply with AC amplitude  $V=2.290$~kV and frequency $\Omega=50$~Hz. For the axial confinement ($Z$-axis), the end-cap electrode was connected to a DC power supply with the DC voltage value up to 300V. Particles were injected from the top of the trap. The AC and DC values remained the same throughout the experiment. After observing a stable configuration, we took a photo of the structure in the XY and XZ-planes (Figure~\ref{fig:greentree}).
The geometrical parameters of the trap are shown in Table~\ref{tab_exp:TrapParameters}. The visualization of the microparticles inside the trap depends on the particle type. Silica non-luminescent particles were visualized using the DPSS-laser ($\lambda_1=532$~nm). Hybrid perovskite nanocomposite PNCs (luminescent particles) were visualized using $\lambda_2=375$~nm laser radiation resulting in photoluminescence (PL) excitation.  The photos of MCCs were obtained with the exposition time (0.5~s) higher than the AC voltage oscillation period (0.02~s). Thus, the photos show the full trajectory track of the particle. 

\begin{table}[ht!]
    \centering
    \begin{tabular}{|c|c|c|c|c|c|c|c|c|c|}
         \hline
         Parameter & End-cap radius & Trap radius & Linear radius & AC amplitude & AC frequency & DC value  \\
         \hline
         Value & 3 mm & 6 mm & 3 mm & 2.290 V & 50 Hz & 300 V\\
         \hline
    \end{tabular}
    \caption{Main characteristics of the experimental setup}
    \label{tab_exp:TrapParameters}
\end{table}

To study the internal structure, we trapped an ensemble of silica microparticles with the average size 28 $\mu$m. The microscope image of the particles studied is shown in Figure~\ref{fig:greentree}a. Our previous investigation showed that the same particles can be successfully trapped in a ``cup-trap''~\cite{rybin2023novel}. However, earlier we dealt with single particles and did not consider the MCCs formation. Figure~\ref{fig:greentree}b-c shows the $XZ$ and $XY$ MCCs projections, which correspond to the front and the top view respectively. The front view shows that the obtained MCC has a Christmas tree-like shape with the height-to-base diameter ratio $2.5$. The top view shows that the particles in the center of MCC (``tree-trunk'' particles) have smaller oscillation amplitude, while the needle-based particles oscillate with the amplitude of up to 2~mm. Moreover, the oscillation track form repeats the shape of the hyperbolic field lines. Thus, we validate the theoretically derived prediction of Christmas tree-like MCCs formation. 

\begin{figure}[ht!]
    \centering
    \includegraphics[width=0.9\textwidth]{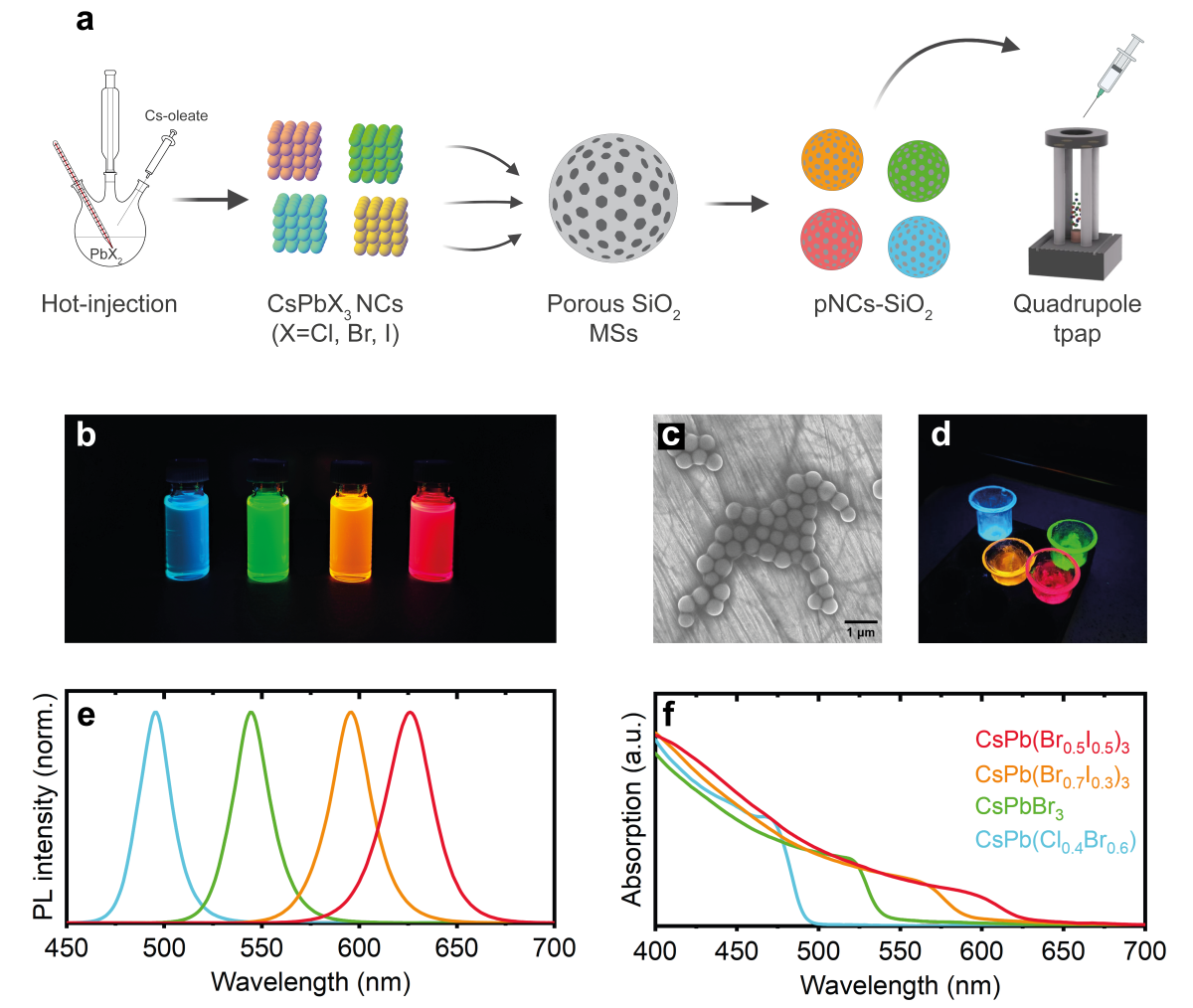}
    \caption{\textbf{Creating a Christmas tree-like Coulomb crystals.} \textbf{a} Experiment scheme. \textbf{b} Colloidal solutions of
CsPbX$_3$ (X = Cl, Br, I) NCs in toluene under UV lamp (\textit{$\lambda$} = 365 nm). \textbf{c} A typical SEM image of porous silica MSs deposited onto Si substrate. \textbf{d} Samples of synthesized pNCs incorporated into MSs under UV lamp (\textit{$\lambda$} = 365 nm). \textbf{e, f} PL (\textit{$\lambda$}\textit{\textsubscript{ex}}= 350 nm) and absorption spectra respectively for pNCs
with different chemical compositions.}
    \label{fig_exp:PerovskiteSpectra}
\end{figure}

So far, we have referred to MCCs in Figure~\ref{fig:sim} as Christmas tree-like. However, it is impossible to imagine Christmas trees without glittering lights. To fix this issues, we used pNCs to create a glittering Christmas tree-like Coulomb crystal. All stages of this part of the experiment are shown step-by-step in Figure~\ref{fig_exp:PerovskiteSpectra}a. At the first stage, we synthesized perovskite nanoparticles with different PL properties. We obtained 4 pNCs samples and labeled them ``blue'', ``green'', ``orange'' and ``red'' according to their PL colour. The photo of as-prepared pNCs solution in toluene under UV lamp is shown in Figure~\ref{fig_exp:PerovskiteSpectra}b. The PL wavelength was tuned using the nanoparticle chemical composition variation. The pNCs synthesis is discussed in the ``Methods'' section.

We then prepared the hybrid structures based on porous silica micropartilces and synthesized pNCs. The SEM image of porous silica micropartilces is shown in Figure~\ref{fig_exp:PerovskiteSpectra}c. The average diameter of a porous silica microsphere is 450~nm. The preparation of pNCs microcomposites (pMSs) is also discussed in the ``Method'' section. Figure~\ref{fig_exp:PerovskiteSpectra}d shows a photo of pMSs powder under UV light. The PL and the absorption spectra of the studied pMSs are shown in Figure~\ref{fig_exp:PerovskiteSpectra}e and f. The composition of the pNCs studied and the corresponding spectral characteristics of pMSs are described in Table~\ref{tab:BGOR}. Finally, the pMSs were used to create glittering multi-colour Christmas tree-like MCCs in the``cup-trap''. 

\begin{figure}[ht!]
    \centering
    \includegraphics[width=1\textwidth]{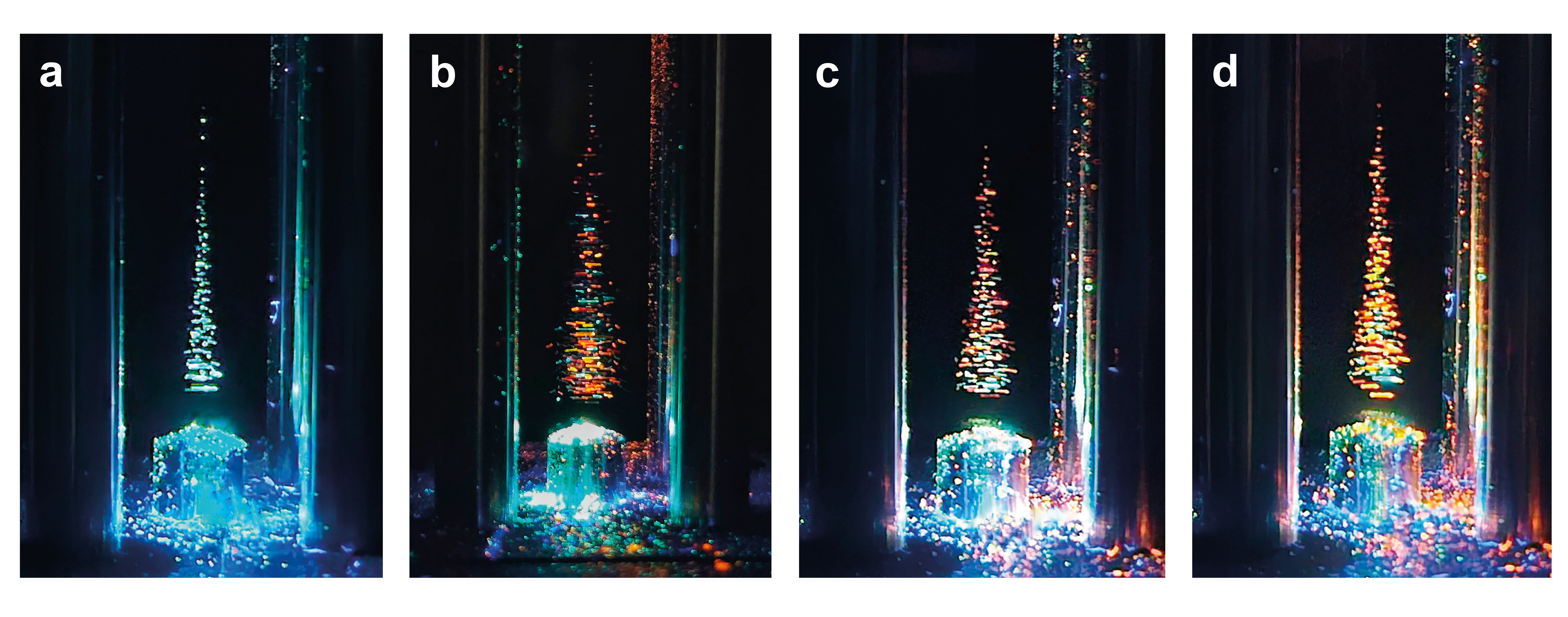}
    \caption{MCCs formation process from the  perovskite nanocomposites particles}    \label{fig_exp:ExperimentalTree}
\end{figure}

To form the glittering multi-colour Christmas tree pMSs, we injected the particles from the top of the trap starting with ``green'' pMSs. As soon as the stable configuration was obtained, we took a photo of the structure in the XZ-plane (Figure~\ref{fig_exp:ExperimentalTree}a). Then we added ``orange'', ``red'' and ``blue'' particles, taking a picture after each step of the ``tree'' formation. The observed CCs are shown in Figure~\ref{fig_exp:ExperimentalTree}. The proposed methods and materials appear to be prospective for creating S\&A objects~(video is available \cite{CTLLink}). 

\section{Conclusion}
We investigated the principles of forming microparticle based Coulomb crystals. For this purpose, we studied the many-body dynamics of the microparticles during their localisation in a vertical quadrupole Paul trap with a single end-cap electrode, a ``cup trap''. We developed a physical model taking into account the electric forces between the linear electrodes and the charged microparticles, the Coulomb interaction between the microparticles, the electrostatic repulsion of the microparticles from the single end-cap electrode and the gravity of microparticles as well as dissipative forces. We showed that microparticles can form stable configurations of MCCs in ``cup traps''. Depending on the gravity contribution, the shape of the MCCs varied from linear chain to cone-shaped 3D MCCs. Moreover, part of the cone-shaped MCCs look like Christmas trees. Our investigation shows that Christmas tree-like MCCs have the following features. Firstly, microparticles located near the trap axis (``tree-trunk'' microparticles) have a smaller oscillation amplitude, while the ``needle-based'' particles oscillate with a higher amplitude. Secondly, the trajectories of the ``needle-based'' particles repeat the shape of hyperbolic field lines. We experimentally verified the formation of Christmas tree-like MCCs in a ``cup-trap'' using silica microspheres. Experimental studies support the predicted features of the MCCs internal structure. Such MCCs may serve as a platform for creating science art objects. Thus, we prepare Christmas tree-like MCCs based on hybrid structures of luminescent perovskite nanoparticles and porous silica microspheres. Our investigation is essentially interdisciplinary including nonlinear dynamics, levitated optomechanics, colloidal chemistry as well as pattern recognition and cognitive neuroscience. We are open to further collaboration with members of various science communities. 

We have prepared our Christmas Tree-like MCCs to open up new horizons in cold plasma physics and in the many-body problem, as well as to show that MCCs can be used as a platform for creating science art objects. We dedicated our S\&A object to the bright holiday of Christmas. However, what makes a simply beautiful object an object of art? Art objects contain complementary embedded meanings. As for our MCCs, Christmas tree is a one of the main symbols of Christmas magic. Thus, every year millions of trees are cut down to decorate our homes and cities, which is the other side of the coin. We would like to highlight the issue of reckless management of natural resources. Let us be honest, the greatest miracle is the world around us. We would like to say: "Save nature and our planet. Save the World. Let's make sure that future generations have their miracles too".

\section{Methods} \label{chapt: Methods}
\subsection{Materials}
1-Octadecene (ODE, 90$\%$, Merk), oleic acid (OA, 90$\%$, Sigma-Aldrich), oleylamine (OlAm, 70$\%$, Sigma-Aldrich), lead (II) iodide (PbI$_2$, 99.999$\%$, Sigma-Aldrich), lead(II) bromide (PbBr$_2$, 99.999$\%$, Sigma-Aldrich), lead (II) chloride (PbCl$_2$, 99.999$\%$, Sigma-Aldrich),  cesium carbonate (Cs$_2$CO$_3$, 99.9$\%$, Sigma-Aldrich), cetyltrimethylammonium bromide (CTAB, $\geq$99$\%$, Acros), tetraethoxysilane (TEOS, $\geq$99$\%$, Acros), ammonium hydroxide solution (NH$_{3}\cdot$ H$_{2}$O, 24 w/w$\%$, $\geq$99.99$\%$, Aldrich), toluene (99.8$\%$, anhydrous, Sigma-Aldrich), ethyl acetate (EtAc, anhydrous, $\geq$99.5$\%$, Sigma-Aldrich). All chemicals were used as purchased.
 
\subsection{Synthesis of CsPbX$_3$ (X = Cl, Br, I) NCs}

To prepare a  Cs-OA solution, 0.814 g of Cs$_2$CO$_3$, 2.5 mL of dried OA, and 30 mL of ODE were added into a 50-mL three-neck flask. The mixture was degassed under vacuum at 120$^\circ C$ for 60 minutes, and then the temperature was increased to 150$^\circ C$ under Ar. After the complete dissolution, the temperature was lowered to 120$^\circ C$ for further use of the mixture.
To synthesize CsPbX$_3$ (X = Cl, Br, I) NCs the reported procedure~\cite{protesescu2015nanocrystals} was used with minor modifications. ODE (5 mL), OA (0.5 mL), OlAm (0.5 mL) and different mixtures of PbX$_2$ (see Table~2) with the total amount 0.188 mmol were loaded into a 25 mL 3-neck flask and dried under vacuum for 1h at 120$^\circ C$. After complete solubilization of the PbX$_2$ salt, the temperature was raised to 180$^\circ C$ under Ar, and the Cs-OA solution (0.4 mL, 0.125 M in ODE, prepared as described above) was quickly injected and after 5s the reaction mixture was cooled by the ice-water bath.
The crude solution was separated by centrifuging at 6000 rpm for 10 min. The precipitant was redissolved in 4 mL of toluene with 8 mL of EtAc added, and centrifuged at 10000 rpm for 10 min. The precipitate was dissolved in toluene and centrifuged at 3000 rpm for 3 minutes. After that the supernatant was removed for further use.

\subsection{Incorporating pNCs in mesoporous microshperes}

\begin{table}[]
\centering
\label{tab:BGOR}
\resizebox{\textwidth}{!}{%
\begin{tabular}{|c|l|c|c|c|c|c|c|c|c|}
\hline
\begin{tabular}[c]{@{}c@{}}Sample \\ name\end{tabular} &
  Chemical formula &
  \begin{tabular}[c]{@{}c@{}}PL peak,\\ nm\end{tabular} &
  \begin{tabular}[c]{@{}c@{}}FWHM,\\ nm\end{tabular} &
  \begin{tabular}[c]{@{}c@{}}$PbX_{2}$, \\ X(1)\end{tabular} &
  \begin{tabular}[c]{@{}c@{}}$PbX_{2}$, \\ X(2)\end{tabular} &
  \begin{tabular}[c]{@{}c@{}}(1) amount \\ of salt, mmol\end{tabular} &
  \begin{tabular}[c]{@{}c@{}}(2) amount \\ of salt, mmol\end{tabular} &
  \begin{tabular}[c]{@{}c@{}}(1) mass \\ of salt, mg\end{tabular} &
  \begin{tabular}[c]{@{}c@{}}(2) mass \\ of salt, mg\end{tabular} \\ \hline
Blue   & CsPb(Cl$_{0.4}$Br$_{0.6}$)$_3$ & 496 & 19 & Cl & Br & 0.113 & 0.075 & 31.42 & 27.6  \\ \hline
Green  & CsPbBr$_{3}$ & 545 & 22 & Cl & Br & 0.188 & - & 69.0 & - \\ \hline
Orange & CsPb(Br$_{0.7}$I$_{0.3}$)$_3$ & 595 & 25 & Br & I  & 0.130 & 0.056 & 47.83 & 25.82 \\ \hline
Red    & CsPb(Br$_{0.5}$I$_{0.5}$)$_3$ & 626 & 27 & Br & I  & 0.093 & 0.093 & 34.13 & 42.87 \\ \hline
\end{tabular}%
}
\caption{
Parameters and composition of pNCs samples}
\end{table}

Spherical microspheres (MSs) of mesoporous silica were synthesized according to the protocols described in ~\cite{stovpiaga2016monodisperse,trofimova2013monodisperse}. Briefly, silica/surfactant (TEOS/CTAB) clusters were meticulously coagulated to create particles with an average pore size of ~3 nm. These particles then underwent controlled annealing at 550$^\circ C$. To produce MSs with an average pore diameter of ~20 nm, a mixture comprising synthesized particles (0.5 g), 1 g of 95.7 vol.$\%$ ethanol, and 4 g of NH$_3 \cdot$H$_2$O (NH$_3$ in $H_{2}O$, 24 wt.$\%$, 99.99$\%$) was placed in an autoclave and maintained at 120$^\circ C$ for 1 hour. The resulting MSs were washed thrice with deionized water, dried at 100$^\circ C$, and stored dry.
Before the NCs incorporation, the MSs were additionally washed with chloroform in an ultrasonic bath to remove residual moisture. After centrifugation, the precipitate was placed into a one-neck flask and dried under vacuum at 250$^\circ C$ for 4 hours. The dried MSs were then placed into a N$_2$-filled glovebox and additionally dried at 200$^\circ C$ for 6 hours. A solution of perovskite nanoparticles was added to the dried MSs and stirred at 400 rpm for 2 hours. The porous microspheres and nanopartilces mixture was centrifuged at 2000 rpm for 10 minutes to separate the MSs from the NCs left in the solution. The precipitate was redispersed in toluene, and the MSs were separated by centrifugation once more to remove any NCs left on the MSs surface. The obtained precipitate was dried under Ar atmosphere for 2 hours to remove the residual solvent for further use.

\subsection{Characterization}
Microscope image of microparticles~(Figure \ref{fig:greentree}a) was taken using a Zeiss LSM-710 confocal microscope. SEM images were taken using a Zeiss Merlin microscope. Absorption spectra were taken using a Shimadzu UV-3600 spectrophotometer. PL spectra in the VIS were taken on a Jasco FP-8200 spectrofluorimeter.

\section{Aknowledgements}
Si-microspheres were fabricated under financial support of the Russian Science Foundation (23-79-00018).
The investigation of the structure by SEM electron microscopy was carried out at the IRC for Nanotechnology of the Science Park of St.Petersburg State University within the framework of project No. AAAA-A19-119091190094.
The investigation of silica microparticles with Zeiss LSM-710 confocal microscope and absorption and PL spectra with spectrophotometer Shimadzu UV-3600 and Jasco FP-8200 spectrofluorimeter were carried out at the R\&EC PhysNano of ITMO University.
The musical accompaniment for the video materials on creating a Christmas tree-like Coulomb crystals was performed by pianist German Markhasin~\cite{CTLLink}.

\section{Competing interests}
The authors declare no competing interests.


\medskip
\bibliographystyle{unsrt}

\end{document}